# Outlier detection and influence diagnostics in network meta-analysis

Hisashi Noma, PhD[*]
Department of Data Science, The Institute of Statistical Mathematics, Tokyo, Japan
ORCID: http://orcid.org/0000-0002-2520-9949

Masahiko Gosho, PhD
Department of Biostatistics, Faculty of Medicine, University of Tsukuba, Tsukuba, Japan

Ryota Ishii, MS
Biostatistics Unit, Clinical and Translational Research Center, Keio University Hospital, Tokyo, Japan

Koji Oba, PhD
Interfaculty Initiative in Information Studies, Graduate School of Interdisciplinary Information Studies, The University of Tokyo, Tokyo, Japan

Toshi A. Furukawa, MD, PhD
Departments of Health Promotion and Human Behavior, Kyoto University Graduate School of Medicine/School of Public Health, Kyoto, Japan

*Corresponding author: Hisashi Noma
 Department of Data Science, The Institute of Statistical Mathematics
 10-3 Midori-cho, Tachikawa, Tokyo 190-8562, Japan
 TEL: +81-50-5533-8440
 e-mail: noma@ism.ac.jp

**Abstract**

Network meta-analysis has been gaining prominence as an evidence synthesis method that enables the comprehensive synthesis and simultaneous comparison of multiple treatments. In many network meta-analyses, some of the constituent studies may have markedly different characteristics from the others, and may be influential enough to change the overall results. The inclusion of these "outlying" studies might lead to biases, yielding misleading results. In this article, we propose effective methods for detecting outlying and influential studies in a frequentist framework. In particular, we propose suitable influence measures for network meta-analysis models that involve missing outcomes and adjust the degree of freedoms appropriately. We propose three influential measures by a leave-one-trial-out cross-validation scheme: (1) comparison-specific studentized residual, (2) relative change measure for covariance matrix of the comparative effectiveness parameters, (3) relative change measure for heterogeneity covariance matrix. We also propose (4) a model-based approach using a likelihood ratio statistic by a mean-shifted outlier detection model. We illustrate the effectiveness of the proposed methods via applications to a network meta-analysis of antihypertensive drugs. Using the four proposed methods, we could detect three potential influential trials involving an obvious outlier that was retracted because of data falsifications. We also demonstrate that the overall results of comparative efficacy estimates and the ranking of drugs were altered by omitting these three influential studies.

Key words: network meta-analysis; contrast-based model; outlier detection; influence diagnostics; multivariate meta-analysis

# 1. Introduction

Network meta-analysis has been gaining prominence as an evidence synthesis method that enables the comprehensive synthesis and simultaneous comparison of multiple treatments. The methodology can synthesize both direct and indirect evidence for all treatment comparisons of interest, and provides estimates of comparative efficacies among treatments, even when no direct comparison evidence exists for some of the included treatments [1,2].

In network meta-analysis, there are often systematic heterogeneities among the synthesized studies, e.g., study designs, participant characteristics, regions, sites, treatment administration, interventions, outcome definitions, etc. Therefore, heterogeneity of the effect sizes is common, and random-effects models have been adopted as standard tools for evidence synthesis [1-4]. However, in many practices of network meta-analysis, some studies might have markedly different characteristics from others, and may exceed the degree of statistical heterogeneity that can be adequately explained by the random-effects model. As in conventional outlier problems in traditional regression analyses [5], these "outlying studies" might lead to biases, potentially yielding misleading results [6]. The evidence obtained from network meta-analyses has already been widely used for public health, clinical practices, health technology assessments, and policy-making, so the identification and influence diagnostics of the outlying studies are relevant issues to prevent serious influences for the various applications.

To detect outliers and assess their influences in network meta-analyses, Zhang et al. [6] recently developed Bayesian influence diagnostic methods for detecting trial-level outliers. Their methods are effectively applicable to Bayesian analyses of arm-based and contrast-based models. However, due to the developments of many useful software packages such as `netmeta`[7] in R and `network`[8] and `network graphs`[9] in Stata, frequentist methods for the contrast-based multivariate random-effects model have been widely adopted in recent practice. These methods do not require complicated computational techniques such as the Markov Chain Monte Carlo (MCMC) or its convergence diagnostics, and are easily tractable for non-statisticians [8,10,11]; thus, they have already been used as alternative standard methods in network meta-analysis. To date, however, there are no effective methods for detecting outliers for use with the frequentist methods of network meta-analysis.

For frequentist inference methods of the conventional univariate meta-analysis, several useful influence diagnosis methods were developed by Viechtbauer and Chueng[12] and Gumedze and Jackson[13]. Also, Negeri and Beyene[14] generalized their methods to the bivariate random-effects model for meta-analyses of diagnostic test accuracy (DTA) studies; but the DTA meta-analyses involve only two outcome variables, and complete observations are usually obtained for both of the two outcome variables, sensitivity and specificity[15]. In network meta-analysis, the dimension of the multivariate model is usually not small (for the illustrative example of antihypertensive drugs[16] in Sections 2 and 6, the dimension was 7), and most of the components of outcome variables for the multivariate model are "missing" (not defined) because most individual trials usually involve only two to five arms.

In this article, we propose effective frequentist methods for identification and influential diagnostics of outlying studies in network meta-analyses. In particular, we propose suitable influence measures for the network meta-analysis models that involve missing outcomes and adjust the degree of freedoms appropriately. We propose three influential measures by a leave-one-trial-out cross-validated scheme: (1) comparison-specific studentized residual, (2) relative change measure for covariance matrix of the comparative effectiveness parameters, (3) relative change measure for heterogeneity covariance matrix. We also propose (4) a model-based approach using a likelihood ratio statistic by a mean-shifted outlier detection model. Note that the proposed methods are generally applicable for multivariate meta-analyses involving missing outcomes[17,18]. We illustrate the effectiveness of the proposed methods by applying them to the network meta-analysis of antihypertensive drugs of Sciarretta et al,[16] which involves an obvious outlying study in the dataset, as will be explained in Section 2. We show that the outlying trial was clearly detectable by all four of the proposed methods; another two trials were also similarly detected. We also demonstrate that the overall results of comparative efficacy estimates and the ranking of drugs were changed by exclusion of the three influential studies.

**2. Motivating example: A network meta-analysis of antihypertensive drugs**

The motivating example is the network meta-analysis of antihypertensive drugs of Sciarretta et al.[16] The authors reviewed 26 randomized controlled trials (223,313 participants) from 1997 to 2009, which compared seven antihypertensive drugs [α-blocker (AB), angiotensin-converting enzyme inhibitor (ACEI), angiotensin II receptor blocker (ARB), β-blocker (BB), calcium channel blocker (CCB), conventional treatment (CT), and diuretic (DD)] and placebo. The efficacy outcome was the incidence of heart failure. Sciarretta et al. assessed the comparative odds ratios (ORs) using network meta-analysis. We present the original dataset in Table 1 and the network plot for this network meta-analysis in Figure 1. Most of the trials were two-arm trials; only two were three-arm trials (STOP-2 and ALLHAT). Note that the Jikei Heart Study was a randomised clinical trial that compared ARB and CT, but the paper describing the main results[19] was retracted from *Lancet* due to misconduct: some falsifications appeared in the published data, and the treatment effect of ARB may have been overestimated. The original network meta-analysis by Sciarretta et al. [16] included this study; consequently, the overall comparative efficacy estimates might have been biased. Therefore, we considered that the Jikei Heart Study might be an influential outlier in this comparative study. However, to date there are no effective methods for assessing the influences of such trials to the overall comparative results by frequentist framework. Thus, we aimed to develop effective methods to provide useful insights for critical evaluations and appropriate interpretations of network meta-analysis in practice.

### 3. Multivariate random-effects model for network meta-analysis

We adopt the contrast-based, multivariate random-effects model for network meta-analysis [8,20]. We consider synthesizing $N$ trials and comparing $p + 1$ treatments. Let $Y_{ij}$ denote an estimator of treatment effect in contrast to a reference treatment (e.g., placebo) for the $j$th treatment in the $i$th trial ($i = 1, 2, \ldots, N; j = 1, 2, \ldots, p$). Commonly used effect measures include mean difference, standardized mean difference, risk difference, risk ratio, odds ratio, and hazard ratio; the ratio measures are usually transformed on a logarithmic scale. For appropriate modelling of within- and between-studies correlations, we consider multivariate outcome variable $\boldsymbol{Y}_i = \left(Y_{i1}, Y_{i2}, \ldots, Y_{ip}\right)^T$. Also, let $\boldsymbol{S}_i$ (a $p \times p$ matrix) be the within-study covariance matrix, which is assumed to be known and fixed to its valid estimate. For the within-study covariance estimators, the Wei and Higgins [21] estimators can be adopted. Note that for trials that do not include a reference treatment, the data augmentation approach of White et al. [3] can be adopted; a quasi-small dataset is added into the reference arm, e.g., 0.001 events for 0.01 patients. Also, most individual clinical trials typically involve only two to five arms, so many of the components of $\boldsymbol{Y}_i$ are usually undefined. We formally regard these undefined components as "missing." The frequentist inferences can be validly implemented for the incomplete outcomes; there have been many theoretical and numerical evidence [20,22].

The multivariate random-effects model for network meta-analysis can be given as

$$Y_i \sim \text{MVN}(\boldsymbol{\theta}_i, \boldsymbol{S}_i), \boldsymbol{\theta}_i \sim \text{MVN}(\boldsymbol{\mu}, \boldsymbol{\Psi}) \qquad (*)$$

$$\boldsymbol{S}_i = \begin{pmatrix} s_{i1}^2 & \rho_{i12}s_{i1}s_{i2} & \cdots & \rho_{i1p}s_{i1}s_{ip} \\ \rho_{i12}s_{i1}s_{i2} & s_{i2}^2 & \cdots & \rho_{i2p}s_{i2}s_{ip} \\ \vdots & \vdots & \ddots & \vdots \\ \rho_{i1p}s_{i1}s_{ip} & \rho_{i2p}s_{i2}s_{ip} & \cdots & s_{ip}^2 \end{pmatrix}$$

$$\boldsymbol{\Psi} = \begin{pmatrix} \tau_1^2 & \kappa_{12}\tau_1\tau_2 & \cdots & \kappa_{1p}\tau_1\tau_p \\ \kappa_{12}\tau_1\tau_2 & \tau_2^2 & \cdots & \kappa_{2p}\tau_2\tau_p \\ \vdots & \vdots & \ddots & \vdots \\ \kappa_{1p}\tau_1\tau_p & \kappa_{2p}\tau_2\tau_p & \cdots & \tau_p^2 \end{pmatrix}$$

where $\boldsymbol{\theta}_i = (\theta_{i1}, \theta_{i2}, \ldots, \theta_{ip})^T$, $\boldsymbol{\mu} = (\mu_1, \ldots, \mu_p)^T$. The grand mean vector $\boldsymbol{\mu}$ expresses the average comparative effectiveness parameters. $\boldsymbol{\Psi}$ is the between-studies covariance matrix. The correlation structure of $\boldsymbol{\Psi}$ can be assumed to be unstructured, but there are rarely enough studies to identify the all variance-covariance parameters. Thus, most network meta-analyses adopt the equal variance assumption $\tau_1^2 = \tau_2^2 = \cdots = \tau_p^2$ [8]; then, all pairwise correlation coefficients $\kappa_{ij}$s should be equal to 0.50 due to the consistency assumption [23,24]. We denote the inverse of the marginal covariance matrix of $Y_i$ as $W_i = (\boldsymbol{\Psi} + \boldsymbol{S}_i)^{-1}$.

For estimation of the model parameters, the log-likelihood function of the multivariate meta-analysis model (*) is given as

$$\ell(\boldsymbol{\mu}, \boldsymbol{\Psi}) = -\frac{1}{2}\sum_{i=1}^{N}\{\log\{\det(\boldsymbol{\Psi} + \boldsymbol{S}_i)\} + (y_i - \boldsymbol{\mu})^T W_i(y_i - \boldsymbol{\mu}) + p_i \log 2\pi\}$$

where $p_i$ is the number of observed outcomes in $Y_i$. Also, $Y_i$ and $S_i$ involve missing components, as mentioned above. In the log-likelihood function, these are shrunk to the sub-vector and sub-matrix for the observed components. In addition, another standard efficient estimation method is the restricted maximum likelihood (REML) estimation. The log-restricted likelihood function is written as

$$\ell_{RL}(\boldsymbol{\mu}, \boldsymbol{\Psi}) = const. + \ell(\boldsymbol{\mu}, \boldsymbol{\eta}) - \frac{1}{2}\log\left\{\det\left(\sum_{i=1}^{N} \boldsymbol{W}_i\right)\right\}$$

The REML estimators of $\{\hat{\boldsymbol{\mu}}, \hat{\boldsymbol{\Psi}}\}$ are obtained by maximizing $\ell_{RL}(\boldsymbol{\mu}, \boldsymbol{\Psi})$. The covariance matrix estimator of $\hat{\boldsymbol{\mu}}$ is given as $V[\hat{\boldsymbol{\mu}}] = \left(\sum_{i=1}^{N} \widehat{\boldsymbol{W}}_i\right)^{-1}$, where $\widehat{\boldsymbol{W}}_i = \left(\widehat{\boldsymbol{\Psi}} + \boldsymbol{S}_i\right)^{-1}$. Because it is well known that the REML estimator has favorable properties [25,26], the REML method is generally adopted in practice.

## 4. Leave-one-trial-out cross-validated influential measures

### *4.1 Comparison-specific studentized residual*

To detect outlying trials and their influence diagnostics, we first discuss leave-one-trial-out cross-validated (LOTOCV)-type influential measures, which are generalized versions of similar measures for conventional univariate meta-analysis proposed by Viechtbauer and Cheung [12].

Straightforward approaches for assessing influence include residual-based methods, which have been extensively discussed in regression diagnosis [5]. For the multivariate meta-analysis model (*), we can define the studentized residuals by a straightforward generalization of the univariate version[12] as

$$\boldsymbol{R}_i = V[\boldsymbol{Y}_i - \hat{\boldsymbol{\mu}}]^{-\frac{1}{2}}(\boldsymbol{Y}_i - \hat{\boldsymbol{\mu}})$$

where $V[\boldsymbol{Y}_i - \hat{\boldsymbol{\mu}}] = \widehat{\boldsymbol{W}}_i^{-1} - \left(\sum_{i=1}^{N} \widehat{\boldsymbol{W}}_i\right)^{-1}$ $(i = 1,2,...,N)$. This multivariate studentized residuals were also proposed by Negeri and Beyene [14] for the bivariate random-effects model of DTA meta-analysis. However, as noted above, there are not missing outcomes in DTA meta-analysis, whereas network meta-analysis models usually involve many missing outcomes. Thus, the dimensions of the multivariate studentized residuals usually differ among the $N$ trials. Also, their individual components might not be intuitively interpreted as influences to the effect-size estimation, as the multivariate measure is studentized as a $p_i$ dimensional vector; the individual components are interpreted solely as their components. Therefore, we propose to use a comparison-specific studentized residual for the influence diagnosis in network meta-analysis:

$$\phi_{ij} = \frac{Y_{ij} - \hat{\mu}_j}{\sqrt{V[Y_{ij} - \hat{\mu}_j]}}$$

where $\hat{\mu}_j$ is the $j$th component of $\hat{\boldsymbol{\mu}}$ ( $i = 1,2,\dots,N; j = 1,2,\dots,p$ ). $V[Y_{ij} - \hat{\mu}_j]$ accords to the ($j$, $j$)-component of $V[\boldsymbol{Y_i} - \hat{\boldsymbol{\mu}}]$. The comparison-specific studentized residual is interpreted as a standardized residual of the effect-size estimator, and is comparable among all pairwise treatment comparisons and all $N$ trials, which usually involve different combinations of treatments. Note that if the $i$th trial does not involve the reference arm, $Y_{ij}$ is defined as a contrast estimate between the $j$th treatment and data-augmented reference treatment. For this case, the standard error (SE) estimate of the residual $Y_{ij} - \hat{\mu}_j$ can be unstable. Therefore, we recommend changing the reference treatment to one of the involved treatment arms for the $i$th trial. Then, $Y_{ij}$ corresponds to a head-to-head comparison estimator, and the comparison-specific studentized residual $\phi_{ij}$ is intuitively interpreted as a residual measure of the direct comparison estimator.

In addition, as is well known in regression diagnosis[5], the naïve studentized residual $\phi_{ij}$ assesses the divergence of the $i$th trial from the overall mean estimate $\hat{\mu}_j$, which is estimated using the information of the $i$th trial itself. Thus, the naïve residual $\phi_{ij}$ might not be a suitable measure of influence assessments because it can involve "optimism". Thus, LOTOCV-type measures have been proposed to circumvent the optimism. Let $\hat{\boldsymbol{\mu}}^{(-i)}, \hat{\boldsymbol{\Psi}}^{(-i)}$ be the REML estimates obtained from the multivariate random-effects model (*) by an $N - 1$ trials dataset that excludes the $i$th trial ($i = 1,2,\dots,N$); we also denote $\widehat{\boldsymbol{W}}_k^{(-i)} = \left(\widehat{\boldsymbol{\Psi}}^{(-i)} + \boldsymbol{S}_k\right)^{-1}$ ($k = 1,2,\ldots,N$). Then, the LOTOCV comparison-specific studentized residual is defined as

$$\psi_{ij} = \frac{Y_{ij} - \hat{\mu}_j^{(-i)}}{\sqrt{V[Y_{ij} - \hat{\mu}_j^{(-i)}]}}$$

where $V[Y_{ij} - \hat{\mu}_j^{(-i)}]$ accords to the ($j, j$) component of $V[\boldsymbol{Y}_i - \widehat{\boldsymbol{\mu}}^{(-i)}] = \left(\widehat{\boldsymbol{W}}_i^{(-i)}\right)^{-1} + \left(\sum_{k \neq i} \widehat{\boldsymbol{W}}_k^{(-i)}\right)^{-1}$ ($i = 1,2,\ldots,N; j = 1,2,\ldots,p$). Note that $\widehat{\boldsymbol{\mu}}^{(-i)}$ and $\widehat{\boldsymbol{\Psi}}^{(-i)}$ are estimated by the $N - 1$ trials dataset that excludes the $i$th trial; thus, $\psi_{ij}$ is interpreted a predicted studentized residual of $j$th component of $i$th trial from the estimated multivariate random-effects model (*) by the other $N - 1$ trials.

When the assumed model (*) is correct, the studentized residual $\psi_{ij}$ follows N(0,1). Thus, a naïve criterion to assess for outliers is comparing 1.96 with the absolute value of $\psi_{ij}$. If the criterion is fulfilled, the corresponding comparison can be considered as a potential outlier. Also, an alternative more precise evaluation of its sampling variation can be conducted using parametric bootstrap. The parametric bootstrap algorithm is given as follows.

*Algorithm 1 (Parametric bootstrap for estimating the sampling distribution of $\psi_{ij}$).*

1. For the multivariate random effects model (*), compute the REML estimates of $\{\boldsymbol{\mu}, \boldsymbol{\Psi}\}$.

2. Resample $\boldsymbol{Y}_1^{(b)}, \boldsymbol{Y}_2^{(b)}, \ldots, \boldsymbol{Y}_N^{(b)}$ from the estimated distribution of (*) with the parameters substituted with $\{\widehat{\boldsymbol{\mu}}, \widehat{\boldsymbol{\Psi}}\}$ via parametric bootstrap, $B$ times ($b = 1,2,\ldots,B$).

3. Compute the LOTOCV studentized residuals $\psi_{ij}^{(b)}$ ($i = 1,2,\ldots,N; j = 1,2,\ldots,p$) for the $b$th bootstrap sample $\boldsymbol{Y}_1^{(b)}, \boldsymbol{Y}_2^{(b)}, \ldots, \boldsymbol{Y}_N^{(b)}$; replicate it for all $B$ bootstrap samples.

4. We can obtain the bootstrap estimate of the sampling distribution of $\psi_{ij}$ by the empirical distribution of $\psi_{ij}^{(1)}, \psi_{ij}^{(2)}, \ldots, \psi_{ij}^{(B)}$.

Typically, the 2.5th and 97.5th percentiles of the bootstrap samples $\psi_{ij}^{(1)}, \psi_{ij}^{(2)}, \ldots, \psi_{ij}^{(B)}$ can be used for the critical values.

*4.2 Relative change for the covariance matrix of $\hat{\boldsymbol{\mu}}$*

Viechtbauer and Cheung[12] and Negeri and Beyene[14] proposed another influence measure that assesses the change in the covariance matrix of the grand mean parameter estimates by the LOTOCV framework. They proposed using the ratio of generalized variances of the estimators of $\boldsymbol{\mu}$ for the leave-one-trial-out dataset and the all-trial dataset. It can be applied similarly to the multivariate random-effects model (*) as an influential measure of network meta-analysis,

$$\text{COVRATIO}_i = \frac{\det(V[\hat{\boldsymbol{\mu}}^{(-i)}])}{\det(V[\hat{\boldsymbol{\mu}}])} = \frac{\det\left(\left(\sum_{k\neq i} \widehat{\boldsymbol{W}}_k^{(-i)}\right)^{-1}\right)}{\det\left(\left(\sum_{k=1}^{N} \widehat{\boldsymbol{W}}_k\right)^{-1}\right)}$$

COVRATIO$_i$ has values in $(0, \infty)$. This measure assesses the relative change of the covariance matrix of $\hat{\boldsymbol{\mu}}$ when the *i*th trial is excluded. If COVRATIO$_i$ is nearly 1, the covariance matrix is not so changed and the exclusion of *i*th trial is not so influential. When COVRATIO$_i$ is larger than 1, the exclusion of the *i*th trial gains the variation of the grand mean estimator $\hat{\boldsymbol{\mu}}$. Then, the *i*th trial is influential as a means of gaining precision, but is not usually an outlying trial that deviates from the overall population and may increase the between-studies heterogeneity. On the contrary, when COVRATIO$_i$ is smaller than 1, the inclusion of the *i*th trial gains the variation of the grand mean estimator $\hat{\boldsymbol{\mu}}$, although the sample size is increased. It means the *i*th trial is influential as means of decreasing precision, and it might also gain the between-studies heterogeneity. Then, the *i*th trial might be an outlying trial that is deviated from the overall population. Thus, the trials with the smallest COVRATIO$_i$ may be outlying and influential trials. For the criterion of assessing outliers, we can also use the bootstrap sampling distribution as discussed in Section 4.1. Substituting $\psi_{ij}$ for COVRATIO$_i$ in Algorithm 1, we can obtain the bootstrap estimate of the sampling distribution for COVRATIO$_i$. Typically, the lower 5th percentile of the bootstrap samples can be used for the critical value.

*4.3 Relative change for the estimate of heterogeneity covariance matrix $\widehat{\boldsymbol{\Psi}}$*

Negeri and Beyene[14] proposed an alternative influence measure that directly assesses the change in the estimate of heterogeneity covariance matrix $\widehat{\boldsymbol{\Psi}}$ using the LOTOCV framework. The proposed measure was the ratio of generalized variances of the estimates of $\boldsymbol{\Psi}$ for the leave-one-trial-out dataset and the all trial dataset. It can also be similarly applied to the multivariate random-effects model (*) of network meta-analysis,

$$\text{PSIRATIO}_i = \frac{\det(\widehat{\boldsymbol{\Psi}}^{(-i)})}{\det(\widehat{\boldsymbol{\Psi}})}$$

The PSIRATIO$_i$ also has values in $(0, \infty)$. This measure assesses the relative change of the estimate of heterogeneity covariance matrix $\widehat{\boldsymbol{\Psi}}$ when the $i$th trial is excluded. If it is nearly 1, the estimate of $\boldsymbol{\Psi}$ is not so altered and the exclusion of $i$th trial is not so influential. When PSIRATIO$_i$ is larger than 1, the exclusion of the $i$th trial increases between-studies heterogeneity. Then, the $i$th trial is usually not an outlying trial that deviates from the overall population. Conversely, when PSIRATIO$_i$ is smaller than 1, the inclusion of the $i$th trial increases between-studies heterogeneity. Then, the $i$th trial might be an outlying trial that deviates from the overall population. Therefore, the trials with the smallest PSIRATIO may be outlying and influential trials. For the criterion of assessing outliers, the bootstrap sampling distribution can be also used. Substituting $\psi_{ij}$ for PSIRATIO$_i$ in Algorithm 1, we can obtain the bootstrap estimate of the sampling distribution for PSIRATIO. Typically, the lower 5th percentile of the bootstrap samples can be used for the critical value.

## 5. Model-based evaluation using a mean-shifted outlier model

Gumedze and Jackson[13] proposed a model-based approach to identifying outlying trials in conventional univariate meta-analysis using a random-effects variance shift model. Negeri and Beyene[14] also proposed another mean-shift model for the bivariate model for DTA meta-analysis. Both papers proposed assessing the significances of the alternative models using the likelihood ratio statistics. Here, we propose another degree of freedom–adjusted mean-shifted model to assess outliers for network meta-analysis.

For a network meta-analysis using the multivariate random-effects model (*), we consider the influence diagnostics of the *i*th trial. We assume that the *i*th trial involves $q_i + 1$ arms, and that degree of freedom is $q_i$. When the reference treatment (e.g., placebo) is not involved in the *i*th trial, the length of $Y_i$ is $q_i + 1$, but the data-augmented arm does not involve substantial statistical information. Thus, we suppose letting the reference treatment change to one of the involved arms of the *i*th trial by sorting the treatment indicators for the influence diagnostics; then, $p_i = q_i$. Under the setting, we propose a mean-shifted outlier model that involves a location-shift parameter $\boldsymbol{\eta} = (\eta_1, \eta_2, \ldots, \eta_p)^T$ for assessing the influence of the *i*th trial. $\eta_j$ is a real-valued parameter if the *j*th component of $Y_i$ is observed, and is strictly 0 otherwise; the length of possibly non-zero components for $\boldsymbol{\eta}$ corresponds to $p_i$. After then, we assume a mean-shifted model for the random-effect of *i*th trial $\boldsymbol{\theta}_i$,

$$\boldsymbol{\theta}_i \sim \text{MVN}(\boldsymbol{\mu} + \boldsymbol{\eta}, \boldsymbol{\Psi})$$

Besides, we assume the ordinary random-effects model (*) for the other $N - 1$ trials. This model assumes that the grand mean parameters of the *i*th trial may diverge from those of the other trials. If the mean-shifted model with non-zero $\boldsymbol{\eta}$ is more plausible than the ordinary multivariate meta-analysis model (*), the *i*th trial can be regarded as a potential outlying trial, and can be assessed by a likelihood ratio statistic. The testing

problem is formulated as

$$H_0: \boldsymbol{\eta} = \mathbf{0} \text{ vs. } H_1: \boldsymbol{\eta} \neq \mathbf{0}$$

where the alternative hypothesis means at least one component of $\boldsymbol{\eta}$ is not equal to 0. Under the alternative hypothesis $H_1$, the log-likelihood function is written as

$$\ell_{[i]}(\boldsymbol{\mu}, \boldsymbol{\eta}, \boldsymbol{\Psi}) = -\frac{1}{2} \sum_{k \neq i} \{\log\{\det(\boldsymbol{\Psi} + \boldsymbol{S}_k)\} + (\boldsymbol{y}_k - \boldsymbol{\mu})^T \boldsymbol{W}_k (\boldsymbol{y}_k - \boldsymbol{\mu}) + p_k \log 2\pi\}$$

$$-\frac{1}{2} \{\log\{\det(\boldsymbol{\Psi} + \boldsymbol{S}_k)\} + (\boldsymbol{y}_i - \boldsymbol{\mu} - \boldsymbol{\eta})^T \boldsymbol{W}_i (\boldsymbol{y}_i - \boldsymbol{\mu} - \boldsymbol{\eta}) + p_i \log 2\pi\}$$

Then, the likelihood ratio statistic is given as

$$T_{[i]} = -2\{\ell(\widetilde{\boldsymbol{\mu}}, \widetilde{\boldsymbol{\Psi}}) - \ell_{[i]}(\widetilde{\boldsymbol{\mu}}_{[i]}, \widetilde{\boldsymbol{\eta}}_{[i]}, \widetilde{\boldsymbol{\Psi}}_{[i]})\},$$

where $\{\widetilde{\boldsymbol{\mu}}, \widetilde{\boldsymbol{\Psi}}\}$ is the maximum likelihood (ML) estimate of the null model (*) and $\{\widetilde{\boldsymbol{\mu}}_{[i]}, \widetilde{\boldsymbol{\eta}}_{[i]}, \widetilde{\boldsymbol{\Psi}}_{[i]}\}$ is the ML estimate of the mean-shifted model for the *i*th trial. The likelihood ratio statistic $T_{[i]}$ has a $\chi^2$-distribution with $p_i$ degrees of freedom under the null hypothesis, asymptotically. Thus, an adequate critical value is $\chi^2_{p_i, 0.95}$ that is the 95th percentile of the $\chi^2$-distribution with $p_i$ degrees of freedom for a test with significance level of 5%. However, as discussed in the previous sections, the conventional $\chi^2$ approximation might be inappropriate. An alternative precise approach is the parametric bootstrap.

*Algorithm 2 (Parametric bootstrap for estimating the sampling distribution of $T_{[i]}$).*

1. Compute the ML estimates $\{\widetilde{\boldsymbol{\mu}}, \widetilde{\boldsymbol{\Psi}}\}$ and $\{\widetilde{\boldsymbol{\mu}}_{[i]}, \widetilde{\boldsymbol{\eta}}_{[i]}, \widetilde{\boldsymbol{\Psi}}_{[i]}\}$.
2. Resample $\boldsymbol{Y}_1^{(b)}, \boldsymbol{Y}_2^{(b)}, \ldots, \boldsymbol{Y}_N^{(b)}$ from the estimated null model of (*), for which the parameters are substituted with $\{\widetilde{\boldsymbol{\mu}}, \widetilde{\boldsymbol{\Psi}}\}$ via parametric bootstrap $B$ times ($b = 1,2,\ldots,B$).
3. Compute the ML estimates $\{\widetilde{\boldsymbol{\mu}}^{(b)}, \widetilde{\boldsymbol{\Psi}}^{(b)}\}$ and $\{\widetilde{\boldsymbol{\mu}}_{[i]}^{(b)}, \widetilde{\boldsymbol{\eta}}_{[i]}^{(b)}, \widetilde{\boldsymbol{\Psi}}_{[i]}^{(b)}\}$ for the *b*th bootstrap sample $\boldsymbol{Y}_1^{(b)}, \boldsymbol{Y}_2^{(b)}, \ldots, \boldsymbol{Y}_N^{(b)}$ and compute the likelihood ratio statistic

$$T_{[i]}^{(b)} = -2\left\{\ell(\widetilde{\boldsymbol{\mu}}^{(b)}, \widetilde{\boldsymbol{\Psi}}^{(b)}) - \ell_{[i]}\left(\widetilde{\boldsymbol{\mu}}_{[i]}^{(b)}, \widetilde{\boldsymbol{\eta}}_{[i]}^{(b)}, \widetilde{\boldsymbol{\Psi}}_{[i]}^{(b)}\right)\right\}$$

This should be replicated for all $B$ bootstrap samples.

4. We can obtain the bootstrap estimate of the sampling distribution of $T_{[i]}$ by the empirical distribution of $T_{[i]}^{(1)}, T_{[i]}^{(2)}, \dots, T_{[i]}^{(B)}$.

Typically, the 95th percentile of the bootstrap samples $T_{[i]}^{(1)}, T_{[i]}^{(2)}, \dots, T_{[i]}^{(B)}$ can be used for critical values for a test with a significance level of 5%.

### 6. Applications to a comparative study of antihypertensive drugs

We analyzed the network meta-analysis of antihypertensive drugs described in Section 2 by the proposed methods. We used the multivariate random-effects model (*) with the equal heterogeneity variances assumption ($\tau_1^2 = \tau_2^2 = \cdots = \tau_p^2$). At first, the comparative OR estimates from all 26 trials are presented in Figure 2(a). The reference was set to placebo, the heterogeneity standard deviation (SD) estimate was 0.099, and substantial heterogeneity was observed. The ranking of treatments and comparative OR estimates were almost the same as those of the original Bayesian analysis using an arm-based model by Sciarretta et al. [16]

In Table 2, we present the results of LOTOCV comparison-specific studentized residuals. We show 15 comparisons with the largest absolute studentized residuals. For the bootstrap estimate of sampling distribution of $\psi_{ij}$s, we conducted 2400 resamplings. There were 28 comparisons of 26 trials. The studentized residuals of most of the 28 comparisons were distributed within both of the 95% standard normal probability interval (−1.96, 1.96) and the parametric bootstrap intervals. However, three comparisons exceeded both of these criteria (TRANSCEND: ARB vs. placebo, Jikei Heart Study: ARB vs. CT, and HYVET: DD vs. placebo). As expected, the Jikei Heart Study had the secondary largest studentized residual (2.241) and exceeded the 97.5th percentile of the bootstrap distribution.

We present the results of LOTOCV analyses of $COVRATIO_i$ and $PSIRATIO_i$ in Table 3. We also conducted 2400 resamplings to estimate their bootstrap distributions. The 15 studies with the smallest $COVRATIO_i$ and $PSIRATIO_i$ are presented. The ranking of the sizes of $COVRATIO_i$ and $PSIRATIO_i$ for the top five trials were consistent for the two influential statistics (TRANSCEND, HYVET, Jikei Heart Study, RENRAAL and ANBP2). Especially for TRANSCEND, HYVET and Jikei Heart Study, the $COVRATIO_i$ values exceeded the bootstrap lower 5th percentile, and were also detected as potential outliers by this criterion. Besides, $PSIRATIO_i$ of HYVET, and Jikei Heart Study exceeded the bootstrap lower 5th percentile, but TRANSCEND did not, although the raw value of $PSIRATIO_i$ of TRANSCEND was the smallest among the 26 trials (0.001) and the 5th percentile was close to it (0.000).

In Table 4, we present the results of the model-based outlier evaluation. The 15 studies with the smallest bootstrap p-values are shown. The number of resamplings was 2400. The likelihood ratio statistic, degree of freedom, bootstrap 95th percentile, and p-value are shown. In these results, the three trials (TRANSCEND, Jikei Heart Study, and HYVET) were also significantly detected by the likelihood ratio tests. The bootstrap 95th percentiles did not necessarily accord to the quantile of the $\chi^2$-distribution, but the testing results were consistent for the 26 trials.

As noted in Section 2, the result of the Jikei Heart Study involved falsifications, and the effect of ARB could be overly estimated (ARB vs. CT; OR: 0.522, 95%CI [confidence interval]: 0.281, 0.939). Also, the TRANSCEND compared ARB vs. placebo, and the result of this study indicated no preventive effect of ARB (OR: 1.047, 95%CI: 0.811, 1.352). Besides, the overall OR estimate of ARB compared with placebo was 0.758 (95%CI: 0.642, 0.896). Thus, TRANSCEND would be detected as a potential outlying study. A possible reason that no preventive effect of ARB indicated is non-study blood-pressure-lowering agents were used more frequently in the placebo group than in the ARB group [27]. The effect of ARB in the trial would be counteracted owing the discrepancy of the concomitant agents between the two groups. Further, HYVET compared DD and placebo, and the OR estimate was 0.375 (95%CI: 0.217, 0.626). The overall OR estimate of this pair was 0.600 (95% CI: 0.487, 0.739). Thus, the efficacy of DD was largely estimated in HYVET relative to the other trials. A possible reason of this result is that HYVET was terminated at the interim analysis and the OR might be largely estimated at the time of termination. In addition, the loss to follow-up rate was considerably large in this trial; it might be because the participants were 80 years of age or older. Missing data have seriously compromised inferences in clinical trials [28]. Note that the number of evaluated comparisons was reasonably large, and the multiplicity issue should be considered, but the frequencies of the detected trials would exceed the range of random errors.

As a whole, TRANSCEND, Jikei Heart Study, and HYVET were consistently detected as potentially outlying and influential trials by all four proposed methods. For a sensitivity analysis, we conducted a synthesized analysis using 23 trials that excluded the three outlying trials; Figure 2(b) shows the results. After excluding the three trials, the heterogeneity SD estimate became markedly smaller (0.009). Thus, the SE estimates of the ORs became smaller as a whole, even though three trials (total participants: 12,852) were excluded. Due to the increase in precision, the comparative OR of CCB vs. placebo was changed to 0.840 (95%CI: 0.735, 0.961), and that of AB vs. placebo was changed to 1.234 (95%CI: 1.012, 1.506); significant differences were observed. Also, because Jikei Heart Study was excluded, the difference of ARB and CT got smaller, and the rankings of ARB and CT was reversed; although the difference was quite small. In addition, the OR estimate for ARB vs. placebo became larger as TRANSCEND was excluded. This would also influence the comparative OR estimates of the other treatments, and the differences vs. placebo grew larger overall. Also, because HYVET was excluded, the difference between DD vs. placebo grew a bit larger; however, this effect was canceled by the fact that TRANSCEND was excluded. These results provide different insights into the comparative efficacy of the seven treatments. Concerning the inconsistency on the network, using the node-splitting method by Dias et al.[29], ARB vs. CT and DD vs. placebo were detected significant inconsistent edges for all 26 trials dataset ($p = 0.027, 0.046$), but they were not significantly detected after excluding the 3 trials ($p = 0.313, 0.281$). The overall treatment rankings and comparative efficacy results would be influenced by the potentially outlying studies, and such sensitivity analyses would provide relevant information for the overall interpretations of network meta-analysis.

## 7. Discussion

The evidence obtained from network meta-analyses has been widely applied in public health, clinical practices, health technology assessment, and policy making. If misleading evidence has been produced as a result of including inappropriate, outlying studies, the impact might be enormous. In this article, we proposed effective methods for detecting outlying and influential studies in network meta-analysis. Through an illustrative example of an antihypertensive drug study[16], we clearly showed that the proposed methods can effectively detect outlying studies (especially, Jikei Heart Study), which may induce misleading evidence. In general practice of network meta-analysis, influence diagnosis tools would be effective tools for preventing misleading results and providing precise evidence.

In addition, in this article, we proposed four frequentist methods that are interpreted as generalized versions of existing outlier evaluating methods of conventional univariate meta-analysis [12,13,30]. The Bayesian outlying detection methods proposed by Zhang et al. [6] are effective tools for Bayesian approaches to network meta-analysis, but cannot be straightforwardly applied to frequentist analyses. The developments and disseminations of the frequentist tools [7-9] for network meta-analysis have been increasingly proceeding. They do not require complicated computation (e.g., MCMC and its convergence diagnosis), and should be familiar to non-statisticians. The development of frequentist methods for outlying detection would help practitioners to obtain precise knowledge from network meta-analysis.

Also, as is noted in Sections 3 and 4, the proposed methods can be applied to general multivariate meta-analysis methods involving partially missing outcomes [17,18]. Although we proposed adopting the comparison-specific studentized residuals, there might be some situations in which the multivariate studentized residuals are more useful. This is an issue that should be addressed in future studies. In addition, we proposed to use a likelihood ratio test for the model-based approach. Recently, Noma et al. [20] discussed the possible inadequacy of the conventional likelihood ratio test under certain situations of network meta-analysis, namely, the $\chi^2$ approximation might be violated especially under small $N$. However, they also showed that the parametric bootstrap adjustment can improve the validity of the inference[20]. Due to recent improvements in computational power, the computational costs of bootstrapping are not prohibitive; consequently, the use of bootstrap approaches is recommended.

In conclusion, this study proposes effective tools for assessing the outlying and influential studies in network meta-analyses. As shown in real data applications, there might be influential studies that can provide misleading evidence in practice. The proposed methods might provide meaningful insights to avoid possible invalid evidence. In addition, for future network meta-analyses, these methods might be used as standard methods to provide precise knowledge, at least in sensitivity analyses.

## Data Availability Statement

The network meta-analysis dataset used in Sections 2 and 6 is part of the published data from Sciarretta et al. [16]

## Acknowledgements

This study was supported by Grant-in-Aid for Scientific Research from the Japan Society for the Promotion of Science (Grant numbers: JP19H04074, JP17K19808).

*Med.* 2011;30(20):2481-2498.

18. Mavridis D, Salanti G. A practical introduction to multivariate meta-analysis. *Stat Methods Med Res.* 2013;22(2):133-158.

19. Mochizuki S, Dahlof B, Shimizu M, et al. Valsartan in a Japanese population with hypertension and other cardiovascular disease (Jikei Heart Study): a randomised, open-label, blinded endpoint morbidity-mortality study (retracted). *Lancet.* 2007;369(9571):1431-1439.

20. Noma H, Nagashima K, Maruo K, Gosho M, Furukawa TA. Bartlett-type corrections and bootstrap adjustments of likelihood-based inference methods for network meta-analysis. *Stat Med.* 2018;37(7):1178-1190.

21. Wei Y, Higgins JP. Estimating within-study covariances in multivariate meta-analysis with multiple outcomes. *Stat Med.* 2013;32(7):1191-1205.

22. Noma H, Nagashima K, Furukawa TA. Permutation inference methods for multivariate meta-analysis. *Biometrics.* 2019; DOI: 10.1111/biom.13134.

23. Higgins JP, Whitehead A. Borrowing strength from external trials in a meta-analysis. *Stat Med.* 1996;15(24):2733-2749.

24. Lu G, Ades AE. Modeling between-trial variance structure in mixed treatment comparisons. *Biostatistics.* 2009;10(4):792-805.

25. Verbeke G, Molenberghs G. *Linear Mixed Models for Longitudinal Data.* New York: Springer; 2000.

26. McCulloch CE, Searle SR, Neuhaus JM. *Generalized, Linear, and Mixed Models.* 2nd ed. New York: Wiley; 2008.

27. TRANSCEND Inverstigators. Effects of the angiotensin-receptor blocker telmisartan on cardiovascular events in high-risk patients intolerant to angiotensin-converting enzyme inhibitors: a randomised controlled trial. *Lancet.* 2008;372(9644):1174-1183.
20

**Table 1.** Summary of network meta-analysis of antihypertensive drugs

| ID | Study | Year | Drug 1 | d/n[‡] | Drug 2 | d/n[‡] | Drug 3 | d/n[‡] |
|----|-------|------|--------|--------|--------|--------|--------|--------|
| 1 | Syst-Eur | 1997 | CCB | 37/2398 | Placebo | 49/2297 | — | — |
| 2 | Syst-China | 1998 | CCB | 4/1253 | Placebo | 8/1141 | — | — |
| 3 | UKPDS | 1998 | ACEI | 12/400 | BB | 9/358 | — | — |
| 4 | ABCD | 1998 | ACEI | 5/235 | CCB | 6/235 | — | — |
| 5 | VHAS | 1997 | CCB | 2/707 | DD | 0/707 | — | — |
| 6 | CAPPP | 1999 | ACEI | 75/5492 | CT | 66/5493 | — | — |
| 7 | NICS-EH | 1999 | CCB | 0/204 | DD | 3/210 | — | — |
| 8 | STOP-2 | 1999 | CCB | 186/2196 | CT | 177/2213 | ACEI | 149/2205 |
| 9 | INSIGHT | 2000 | CCB | 26/3157 | DD | 12/3164 | — | — |
| 10 | NORDIL | 2000 | CCB | 63/5410 | CT | 53/5471 | — | — |
| 11 | ALLHAT | 2000 | AB | 491/9067 | DD | 420/15268 | — | — |
| 12 | HOPE | 2000 | ACEI | 417/4645 | Placebo | 535/4652 | — | — |
| 13 | RENRAAL | 2001 | ARB | 89/751 | Placebo | 127/762 | — | — |
| 14 | LIFE | 2002 | ARB | 153/4605 | BB | 161/4588 | — | — |
| 15 | CONVINCE | 2003 | CCB | 126/8179 | CT | 100/8297 | — | — |
| 16 | ALLHAT | 2002 | CCB | 706/9048 | DD | 870/15255 | ACEI | 612/9054 |
| 17 | VALUE | 2004 | ARB | 354/7649 | CCB | 400/7596 | — | — |
| 18 | ANBP2 | 2003 | ACEI | 69/3044 | DD | 78/3039 | — | — |
| 19 | SHELL | 2003 | CCB | 23/942 | DD | 19/940 | — | — |
| 20 | FEVER | 2005 | CCB | 18/4841 | Placebo | 27/4870 | — | — |
| 21 | ASCOT-BPLA | 2005 | CCB | 134/9639 | BB | 159/9618 | — | — |
| 22 | E-COST | 2005 | ARB | 35/1053 | CT | 41/995 | — | — |
| 23 | Jikei Heart Study | 2007 | ARB | 19/1541 | CT | 36/1540 | — | — |
| 24 | HYVET | 2008 | DD | 22/1933 | Placebo | 57/1912 | — | — |
| 25 | ONTARGET | 2008 | ACEI | 514/8576 | ARB | 537/8542 | — | — |
| 26 | TRANSCEND | 2008 | ARB | 134/2954 | Placebo | 129/2972 | — | — |

[†] AB: α-blocker; ACEI: angiotensin-converting enzyme inhibitor; ARB: angiotensin II receptor blocker; BB: β-blocker; CCB: calcium channel blocker; CT: conventional treatment; DD: diuretic

[‡] d: Number of incidences of heart failure; n: no. of participants.

**Table 2**. Results of outlier evaluations using the studentized deleted residual $\psi_{ij}$: 15 comparisons with the largest absolute studentized deleted residuals

| ID | Study | Comparison | $\psi_{ij}$ | Bootstrap lower 2.5th percentile | Bootstrap upper 2.5th percentile |
|---|---|---|---|---|---|
| 26 | TRANSCEND | ARB vs. Placebo | −2.531 | −2.101 | 2.138 |
| 23 | Jikei Heart Study | ARB vs. CT | 2.241 | −2.016 | 1.998 |
| 24 | HYVET | DD vs. Placebo | 2.001 | −2.074 | 1.841 |
| 18 | ANBP2 | ACEI vs. DD | 1.746 | −2.100 | 1.992 |
| 7 | NICS-EH | CCB vs. DD | 1.499 | −1.911 | 2.032 |
| 21 | ASCOT-BPLA | CCB vs. BB | 1.312 | −2.116 | 2.199 |
| 9 | INSIGHT | CCB vs. DD | −1.243 | −1.954 | 1.937 |
| 6 | CAPPP | ACEI vs. CT | −1.193 | −1.959 | 2.060 |
| 15 | CONVINCE | CCB vs. CT | −1.119 | −2.000 | 2.153 |
| 2 | Syst-China | CCB vs. Placebo | 1.013 | −1.890 | 2.017 |
| 14 | LIFE | ARB vs. BB | −0.877 | −2.088 | 2.146 |
| 3 | UKPDS | ACEI vs. BB | −0.877 | −1.896 | 1.929 |
| 22 | E-COST | ARB vs. CT | 0.855 | −2.047 | 1.936 |
| 13 | RENRAAL | ARB vs. Placebo | 0.825 | −1.977 | 2.010 |
| 5 | VHAS | CCB vs. DD | −0.820 | −1.915 | 1.948 |

[†] AB: α-blocker; ACEI: angiotensin-converting enzyme inhibitor; ARB: angiotensin II receptor blocker; BB: β-blocker; CCB: calcium channel blocker; CT: conventional treatment; DD: diuretic

**Table 3**. Results of outlier evaluations using COVRATIO$_i$ and PSIRATIO$_i$: 15 studies with the smallest influential statistics

| | Evaluation by COVRATIO$_i$ | | | | Evaluation by PSIRATIO$_i$ | | |
|---|---|---|---|---|---|---|---|
| ID | Study | COVRATIO$_i$ | Bootstrap lower 5th percentile | ID | Study | PSIRATIO$_i$ | Bootstrap lower 5th percentile |
| 26 | TRANSCEND | 0.067 | 0.133 | 26 | TRANSCEND | 0.001 | 0.000 |
| 24 | HYVET | 0.079 | 0.343 | 24 | HYVET | 0.002 | 0.031 |
| 23 | Jikei Heart Study | 0.222 | 0.515 | 23 | Jikei Heart Study | 0.029 | 0.133 |
| 13 | RENRAAL | 0.650 | 0.239 | 13 | RENRAAL | 0.241 | 0.002 |
| 18 | ANBP2 | 0.683 | 0.212 | 18 | ANBP2 | 0.300 | 0.003 |
| 2 | Syst-China | 0.717 | 0.778 | 21 | ASCOT-BPLA | 0.334 | 0.000 |
| 20 | FEVER | 0.852 | 0.494 | 2 | Syst-China | 0.484 | 0.468 |
| 1 | Syst-Eur | 0.991 | 0.376 | 20 | FEVER | 0.615 | 0.121 |
| 5 | VHAS | 1.005 | 0.918 | 1 | Syst-Eur | 0.733 | 0.035 |
| 7 | NICS-EH | 1.036 | 0.903 | 14 | LIFE | 0.826 | 0.000 |
| 9 | INSIGHT | 1.107 | 0.564 | 5 | VHAS | 1.004 | 0.777 |
| 4 | ABCD | 1.164 | 0.808 | 7 | NICS-EH | 1.066 | 0.767 |
| 3 | UKPDS | 1.195 | 0.735 | 9 | INSIGHT | 1.092 | 0.167 |
| 21 | ASCOT-BPLA | 1.242 | 0.232 | 15 | CONVINCE | 1.219 | 0.004 |
| 22 | E-COST | 1.353 | 0.502 | 3 | UKPDS | 1.238 | 0.387 |

**Table 4.** Results of outlier evaluations by the likelihood ratio statistic: 15 studies with the smallest bootstrap p-values

| ID | Study | LR statistic | df[‡] | Bootstrap 95th percentile | Bootstrap p-value |
|---|---|---|---|---|---|
| 26 | TRANSCEND | 6.801 | 1 | 3.930 | 0.012 |
| 23 | Jikei Heart Study | 5.282 | 1 | 3.916 | 0.023 |
| 24 | HYVET | 4.326 | 1 | 3.776 | 0.036 |
| 18 | ANBP2 | 3.503 | 1 | 4.040 | 0.068 |
| 21 | ASCOT-BPLA | 2.467 | 1 | 3.904 | 0.118 |
| 7 | NICS-EH | 2.247 | 1 | 3.844 | 0.129 |
| 15 | CONVINCE | 1.768 | 1 | 3.845 | 0.193 |
| 6 | CAPPP | 1.706 | 1 | 3.798 | 0.202 |
| 9 | INSIGHT | 1.619 | 1 | 3.864 | 0.213 |
| 14 | LIFE | 1.524 | 1 | 3.785 | 0.215 |
| 13 | RENRAAL | 1.279 | 1 | 3.847 | 0.263 |
| 2 | Syst-China | 1.149 | 1 | 3.922 | 0.288 |
| 22 | E-COST | 0.867 | 1 | 3.965 | 0.351 |
| 1 | Syst-Eur | 0.845 | 1 | 3.801 | 0.357 |
| 3 | UKPDS | 0.805 | 1 | 3.562 | 0.363 |

[‡] df: degree of freedom.

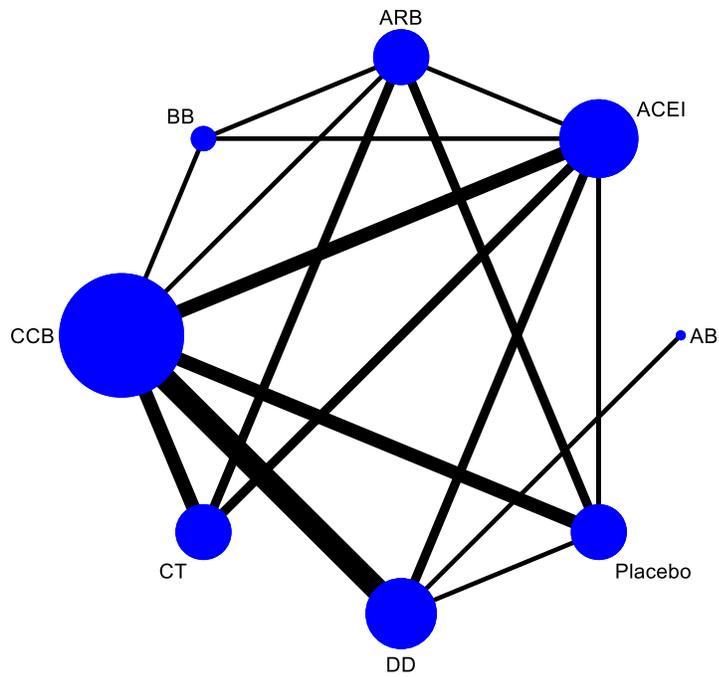

**Figure 1.** Network plot for the network meta-analysis of 26 clinical trials (AB: α-blocker; ACEI: angiotensin-converting enzyme inhibitor; ARB: angiotensin II receptor blocker; BB: β-blocker; CCB: calcium channel blocker; CT: conventional treatment; DD: diuretic).

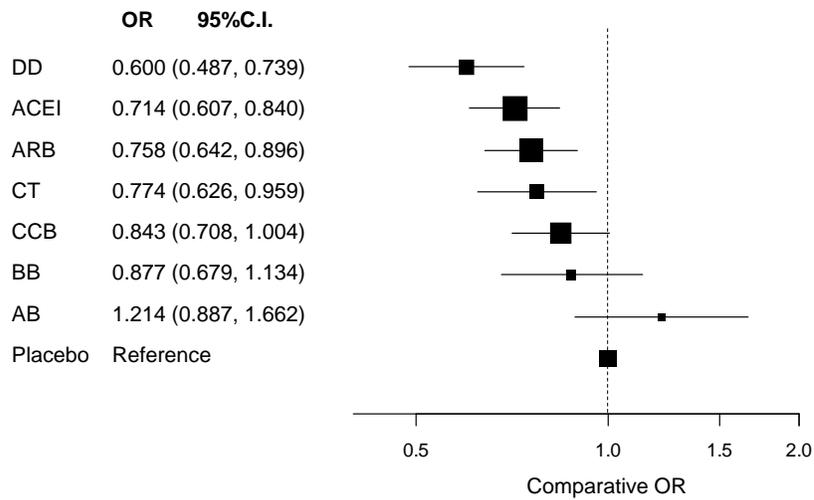

(a) Network meta-analysis by all 26 trials

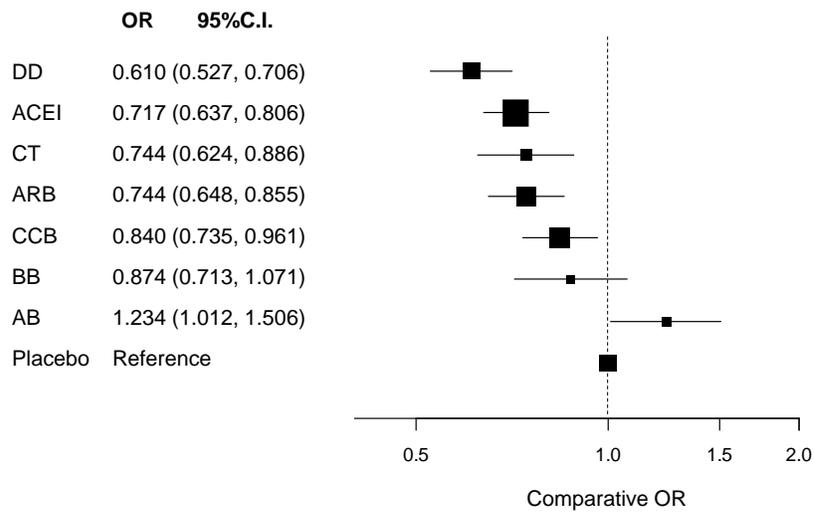

(b) Sensitivity analysis excluding TRANSCEND, Jikei Heart Study and HYVET

**Figure 2.** Confidence interval plots for the comparative odds ratio (OR) estimates: (a) by all 26 trials; (b) sensitivity analysis excluding TRANSCEND, Jikei Heart Study, and HYVET (AB: α-blocker; ACEI: angiotensin-converting enzyme inhibitor; ARB: angiotensin II receptor blocker; BB: β-blocker; CCB: calcium channel blocker; CT: conventional treatment; DD: diuretic).